%% file: ICASSP_main.tex
\documentclass{article}
\usepackage{cite}
\usepackage{spconf,algorithm,algorithmic,amsmath,amssymb,amsthm,bbm,color,graphicx,microtype,url}
\usepackage{subcaption}

\input{./Definitions.tex}

\title{Near-field Localization with 1-bit Quantized Hybrid A/D Reception }
\name{Ioannis Gavras,$^1$ Italo Atzeni,$^2$ and George C. Alexandropoulos$^1$
\thanks{The work of I. Gavras and G. C. Alexandropoulos was supported by the Smart Networks and Services Joint Undertaking (SNS JU) project TERRAMETA under the European Union's Horizon Europe research and innovation programme under Grant Agreement no. 101097101, including top-up funding by UK Research and Innovation (UKRI) under the UK government's Horizon Europe funding guarantee. The work of I.~Atzeni was supported by the Research Council of Finland (336449 Profi6, 346208 6G~Flagship, 348396 HIGH-6G, and 357504 EETCAMD).}}
\address{$^1$Dept. of Informatics and Telecommunications, National and Kapodistrian University of Athens, Greece\\
$^2$Centre for Wireless Communications, University of Oulu, Finland}

\begin{document}
\maketitle

\begin{abstract}
In this paper, we consider a hybrid Analog and Digital (A/D) receiver architecture with an extremely large Dynamic Metasurface Antenna (DMA) and an $1$-bit resolution Analog-to-Digital Converter (ADC) at each of its reception radio-frequency chains, and present a localization approach for User Equipment (UE) lying in its near-field regime. The proposed algorithm scans the UE area of interest to identify the DMA-based analog combining configuration resulting to the peak in a received pseudo-spectrum, yielding the UE position estimation in three dimensions. Our simulation results demonstrate the validity of the proposed scheme, especially for increasing DMA sizes, and showcase the interplay among various system parameters.
\end{abstract}


\section{Introduction}
Extremely large antenna arrays~\cite{xlmimo} and ultra-large bandwidths at millimeter wave, and beyond, spectra~\cite{TRmag} constitute respectively a promising technology and core feauture of upcoming sixth Generation (6G) wireless networks, contributing in boosting communication data rates, energy efficiency, and sensing resolution~\cite{risisac}. To this end, holographic Multiple-Input and Multiple-Output (MIMO) transceivers, comprising sub-wavelength-spaced elements in almost continuous antenna apertures, are lately receiving substantial research and development interest~\cite{HMIMO_survey}. One of the promising hybrid Analog and Digital (A/D) HMIMO technologies are Dynamic Metasurface Antennas (DMAs), consisting of microstrips of metamaterial collections with tunable responses~\cite{Shlezinger2021Dynamic,Nlos_DMA,zhang2022beam,yang2023near,xu2021dynamic,lwei2024tri}.

The flexibility of DMAs in shaping Radio Frequency (RF) waves in the analog domain is recently exploited for different applications in (sub-)THz frequencies, leveraging the short wavelengths and large bandwidths to achieve remarkable spatiotemporal resolution~\cite{TRmag,zhang2022beam,qlo2023ultra,gca2024risthz}. To this end, there has been growing interest for efficient DMA hardware designs~\cite{qlo2023ultra,xma2024bit} and fabrication schemes~\cite{adp2024adapt}.

Extremely large DMAs at high frequencies usually result in wireless system deployments in the near-field regime \cite{gavras2023full,gavras2024track}, where precise control over the spatial characteristics of the transmitted and received signals becomes crucial for accurate communications, localization, and sensing. However, to the best of the authors' knowledge, localization schemes with hardware-efficient DMA structures have not been studied. To fill this gap, in this paper, we capitalize on a recent DMA Receiver (RX) architecture, comprising $1$-bit resolution Analog-to-Digital Converters (ADCs) at each of its reception RF chains~\cite{pga2024meta} attached to a disctinct group of metamaterials, and present a localization approach for single-antenna User Equipment (UE) lying in its near-field regime. The proposed approach is tailored to the hardware constraints of the DMA-based hybrid A/D RX and deploys a simple thresholding mechanism for accurate UE position parameter estimation. The presented simulation results for a sub-THz frequency range substantiate the effectiveness of the proposed localization scheme, which is showcased to compensate the quantization loss induced by the $1$-bit ADCs with increasing numbers of metamaterials at the DMA, outperforming a state-of-the-art benchmark relying on full resolution ADCs. 

\textit{Notations:} Vectors and matrices are denoted by boldface lowercase and boldface capital letters, respectively. The Hermitian transpose of $\mathbf{A}$ is denoted by $\mathbf{A}^{\rm H}$, $[\mathbf{A}]_{i,j}$ is the $(i,j)$th element of $\mathbf{A}$, $\|\mathbf{A}\|$ returns $\mathbf{A}$'s Euclidean norm, and $|a|$ is the amplitude of a complex scalar $a$. $\mathbb{C}$ is the complex number set, $\mathcal{S}$ returns the cardinality of set $\mathcal{S}$, and $\jmath$ is the imaginary unit. $\mathbb{E}\{\cdot\}$ is the expectation operator and $\mathbf{x}\sim\mathcal{CN}(\mathbf{a},\mathbf{A})$ indicates a complex Gaussian random vector with mean $\mathbf{a}$ and covariance matrix $\mathbf{A}$.

\section{System and Signal Models}\label{sec: system_signal}
\begin{figure}
		\centering
		\includegraphics[width=0.95\linewidth]{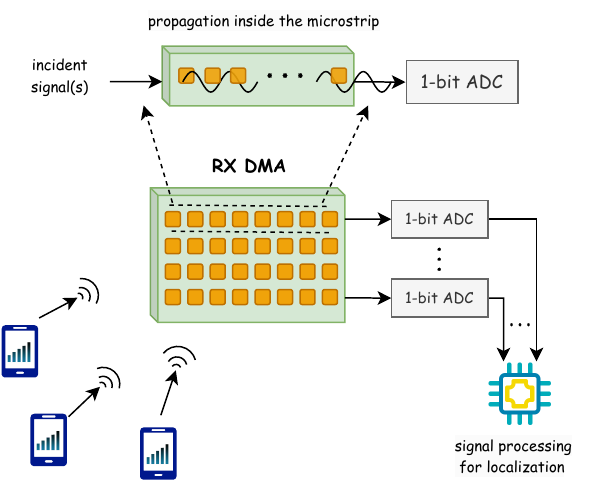}
		\caption{The proposed DMA-based hybrid A/D RX architecture with an $1$-bit resolution ADC at each reception RF chain.} 
  \label{fig:LWA}
\end{figure}
We consider a DMA-based RX with an extremely large number of metamaterials grouped in microstrips~\cite{Shlezinger2021Dynamic}, with each microstrip attached to a reception RF chain including an $1$-bit resolution ADC, as presented in~\cite{pga2024meta} and illustrated in Fig.~\ref{fig:LWA}. This hybrid A/D reception system, possibly integrating sub-wavelength-spaced metamaterials within its aperture~\cite{HMIMO_survey}, wishes to localize a single-antenna UE lying in its near-field region via appropriate optimization of the metamaterials' phase profiles and processing of the baseband received signal. 

We assume that the RX DMA panel is placed on the positive $xz$-plane with its first microstrip lying at the origin. It comprises $N_{\rm RF}$ microstrips each consisting of $N_{\rm E}$ metamaterials. This arrangement results in a total of $N \triangleq N_{\rm RF}N_{\rm E}$ metamaterials, acting as an extremely large planar antenna array. To this end, the distance between any pair of microstrips is represented by $d_{\rm RF}$, while the inter-element distance within each microstrip is $d_{\rm E}$. We finally assume for simplicity that the UE is positioned along the $z$-axis.

We define the $N\times N$ diagonal matrix $\P_{\rm RX}$ with each non-zero elements modeling the signal propagation inside the DMA microstrips. In particular, $\forall$$i=1,2,\ldots,N_{\rm RF}$ and $\forall$$n = 1,2,\ldots,N_{\rm E}$ holds for this matrix that~\cite{Xu_DMA_2022}:
\begin{align}\label{eq: RX_Sig_Prop}
    [\P_{\rm RX}]_{(i-1)N_{\rm E}+n,(i-1)N_{\rm E}+n}\! \triangleq \!\exp{(-\rho_{i,n}(\alpha_i + \jmath\beta_i))},
\end{align}
where, $\alpha_i$ represents the waveguide attenuation coefficient, $\beta_i$ corresponds to the wavenumber, and $\rho_{i,n}$ indicates the position of the $n$th element in the $i$th microstrip. Let $w^{\rm RX}_{i,n}$ represent the adaptable response (i.e., analog weight) of the $n$th metamaterial in the $i$th microstrip. These weights adhere to a Lorentzian-constrained phase model and belong to the phase profile codebook $\mathcal{W}$, as follows:
\begin{align}
    w^{\rm RX}_{i,n} \in \mathcal{W}\triangleq \left\{\frac{\jmath+e^{\jmath\phi}}{2}\Big|\phi\in\left[-\frac{\pi}{2},\frac{\pi}{2}\right]\right\}.
\end{align}
Hence, the analog RX combining matrix $\W_{\rm RX}\in\mathbb{C}^{N\times N_{\rm RF}}$ is given by:
\begin{align}
    [\W_{\rm RX}]_{(i-1)N_{\rm E}+n,j} = \begin{cases}
    w^{\rm RX}_{i,n},&  i=j\\
    0,              & i\neq j
\end{cases}.
\end{align}


\subsection{Channel Model}
We investigate wireless operations in the (sub-)THz frequency band and focus specifically in an near-field signal propagation environment. To this end, the complex-valued $1\times N$ channel matrix between the DMA-based RX and the single-antenna UE is modeled as follows:
\begin{align}
    \label{eqn:UL_chan}
    [\h]_{(i-1)N_{\rm E}+n} \triangleq \alpha_{i,n} \exp\left(\frac{\jmath2\pi}{\lambda} r_{i,n}\right),
\end{align}
where $r_{i,n}$ denotes the distance between the UE's antenna and the $n$th reception meta-element of the $i$th microstrip. Additionally, $\alpha_{i,n}$ represents the attenuation factor including the molecular absorption coefficient $\kappa_{\rm abs}$ at (sub-)THz frequencies and defined as:
\begin{align}\label{eq: atn}
    \alpha_{i,n} \triangleq \sqrt{F(\theta_{i,n})} \frac{\lambda}{4\pi r_{i,n}} \exp\left(-\frac{\kappa_{\rm abs}r_{i,n}}{2}\right)
\end{align}
with $\lambda$ being the signal wavelength, while $F(\cdot)$ represents each metamaterial's radiation profile. This profile is modeled for an elevation angle $\theta$ as follows:
\begin{align}
    F (\theta) = \begin{cases}
    2(b+1)\cos^{b}(\theta),& {\rm if}\, \theta\in[-\frac{\pi}{2},\frac{\pi}{2}]\\
    0,              & {\rm otherwise}
\end{cases}.
\end{align}
In the latter expression, $b$ determines the boresight antenna gain which depends on the specific DMA technology. 

The spherical coordinates of the UE's antenna, situated in the near-field of the RX DMA, are denoted as $(r, \theta, \varphi)$ with elements representing respectively the distance from the origin, elevation and azimuth angles. Each distance $r_{i,n}$ in \eqref{eqn:UL_chan} and \eqref{eq: atn} can be computed as:
\begin{align}\label{eq: dist}
    \nonumber &r_{i,n}\! =\! \!\Big(\!(r\sin\theta\cos\varphi -(i\!-\!1)d_{\rm RF})^2 +\\ &(r\sin\theta\sin\varphi)^2 + (r\cos\theta\!-\!(n\!-\!1)d_{\rm E})^2\Big)^{\frac{1}{2}},
\end{align}
Note that the elevation angle of the UE's antenna with respect to the $n$th reception meta-element of each $i$th microstrip is given by the following expression:
\begin{align}\label{eq:thetas}
    \theta_{i,n} \triangleq \sin^{-1}\left({\frac{|(n-1)d_{\rm E}-r\cos{\theta}|}{r_{i,n}}}\right).
\end{align}

\subsection{Received Signal Model}
The baseband received signal $\y\in\Compl^{N_{\rm RF}\times1}$ at the output of the RX RF chains can be mathematically expressed as follows:
\begin{align}
    \y \triangleq \W_{\rm RX}^{\rm H}\P_{\rm RX}^{\rm H}\h^{\rm H}s + \W_{\rm RX}^{\rm H}\P_{\rm RX}^{\rm H}\n,
\end{align}
where $s\in\Compl$ (in practice, it belong in a finite discrete modulation set) indicates the transmitted pilot symbol from the UE and $\n\sim\mathcal{CN}(\mathbf{0},\sigma^2\mathbf{I}_N)$ denotes the Additive White Gaussian Noise (AWGN) vector due to thermal noise. It is further assumed that, in each Transmission Time Interval (TTI), the UE's pilot symbol is power limited such that $\mathbb{E}\{|s|^2\}\leq P_{\rm max}$  with $P_{\rm max}$ representing the maximum transmission power. As illustrated in Fig.~\ref{fig:LWA}, each RX RF chain comprises an $1$-bit resolution ADC. Following \cite{abdelhameed2023enhanced}, the quantization procedure for each $i$th microstrip output (i.e., each $i$th element of $\y$) can be represented via the following quantization function:
\begin{align}\label{eq: quan}
    \mathcal{Q}([\y]_i)\!=\!\begin{cases}
    0.5\!\left(\text{sign}\left(\mathcal{R}\{[\y]_i\}\right)\!+\!\jmath\text{sign}\left(\mathcal{I}\{[\y]_i\}\right)\right)\!,& \!\! \!\! \!\! [\g]_i\!\leq\! \gamma_q\\
    0,               & \!\! \!\! \!\! \text{otherwise}
    \end{cases},
\end{align}
where $[\g]_i\triangleq|[\y]_i-k_i|$ with $k_i$ being a Dynamic Direct Current (DC) offset. Note that, according to this formula, only when the amplitude of the received signal at each $i$th microstrip, minus the DC offset, falls below a predefined threshold $\gamma_q$, it undergoes $1$-bit quantization. Otherwise, the signal is treated as environmental noise, providing a zero at the output of the quantizer. We denote the $N_{\rm RF}$-element received signal vector after quantization as $\y_{q} \triangleq \mathcal{Q}(\y)$. Capitalizing on our near-field channel model, we can derive a closed-form formula for the DC offset and the quantization threshold as follows:
\begin{align}\label{eq: thresh}
    &k \triangleq \W_{\rm RX}^{\rm H}\P_{\rm RX}^{\rm H}\widehat{\h}^{\rm H},\quad\gamma_q \triangleq \sqrt{N\sigma^2}.
\end{align}
In the former expression, the DC offset $k$ aims to replicate the gain of the received pilot signal, by utilizing the reconstructed uplink channel $\widehat{\h}$ via \eqref{eqn:UL_chan} using a priori knowledge of the UE 3D position coordinates $\widehat{r}$, $\widehat{\theta}$, and $\widehat{\varphi}$. Through this operation, our objective is to neutralize the impact of the true signal, leaving only the processed AWGN. The power of the processed AWGN must be less than or equal to the value $\gamma_q$, where $\gamma_q$ is equivalent to the Root Mean Square Error (RMSE) value of the expected power of the AWGN vector (i.e., $\mathbb{E}\{\|\n\|\}$), ensuring conditions suitable for the quantization process.

\section{Proposed Near-Field Localization}
In this section, we optimize the free parameters of the considered DMA-based hybrid A/D RX architecture with $1$-bit resolution ADCs for near-field localization. 

\subsection{Analog Combining Optimization}
To achieve precise UE localization, we focus on maximizing the received Signal-to-Noise Ratio (SNR) in the uplink direction through the optimization of the RX analog combining matrix. This objective can be expressed mathematically as:
\begin{align}
        \mathcal{OP}&:\nonumber\underset{\substack{\widetilde{\W}_{\rm RX}}}{\max} \quad  \|\widetilde{\W}_{\rm RX}^{\rm H}\P_{\rm RX}^{\rm H}\widehat{\h}\|^2\,\,\text{\text{s}.\text{t}.}\, 
        \, \widetilde{w}^{\rm RX}_{i,n} \in \mathcal{W},\nonumber
\end{align}
For a given UE coordinate tuple $(\widehat{r},\widehat{\theta},\widehat{\varphi})$, the near-field channel gain vector $\widehat{\h}$ can be constructed via \eqref{eqn:UL_chan}. To solve $\mathcal{OP}$, we first constrain its $\widetilde{\W}_{\rm RX}$ elements in the set $\mathcal{F}\in\{e^{j\phi}|\phi\in\left[-\pi/2,\pi/2\right]\}$ (e.g., a Discrete Fourier Transform codebook) having constant amplitude and variable phase values. Subsequently, we perform an 1D search. Finally, given $\widetilde{\W}_{\rm RX}$ and accounting for the signal propagation inside the microstrips, the DMA analog combining weights are obtained as:
\begin{align}\label{eq:weights}
  w^{\rm RX}_{i,n} \triangleq 0.5\left(\jmath+\widetilde{w}^{\rm RX}_{i,n}e^{\jmath\rho_{i,n} \beta_{i}}\right).
\end{align}

\begin{algorithm}[!t]
    \caption{Estimation of UE Position Coordinates}
    \label{alg:loc}
    \begin{algorithmic}[1]
        \renewcommand{\algorithmicrequire}{\textbf{Input:}}
        \renewcommand{\algorithmicensure}{\textbf{Output:}}
        \REQUIRE $\mathcal{S}, \gamma_q$, and $P_{\rm max}$. 
        \ENSURE $(\widehat{r},\widehat{\theta},\widehat{\varphi})$. 
        \STATE Initialize $\mathcal{P}(\widehat{r}_p,\widehat{\theta}_p,\widehat{\varphi}_p)= 0$ $\forall (\widehat{r}_p,\widehat{\theta}_p,\widehat{\varphi}_p)\in\mathcal{S}$.
        \FOR{every $(\widehat{r}_p,\widehat{\theta}_p,\widehat{\varphi}_p) \in \mathcal{S}$ }
                \STATE Construct the virtual channel $\widehat{\h}$ using $(\widehat{r}_p,\widehat{\theta}_p,\widehat{\varphi}_p)$ in \eqref{eqn:UL_chan}, and update $k$ via \eqref{eq: thresh}.
                \STATE Substitute $\widehat{\h}$ into $\mathcal{OP}$ and solve for $\W_{\rm RX}$.
                \STATE Apply $\W_{\rm RX}$ and obtain $\y_q$ resulting from the quantization in \eqref{eq: quan}. 
                \STATE Set $\mathcal{P}(\widehat{r}_p,\widehat{\theta}_p,\widehat{\varphi}_p) = \|\y_q\|^2$. 
        \ENDFOR
        \STATE Conduct grid search in the pseudo-spectrum $\mathcal{P}$ to find the peak corresponding to $(\widehat{r},\widehat{\theta},\widehat{\varphi})$.
    \end{algorithmic}
\end{algorithm}
\subsection{UE Coordinates' Estimation}
We present an on-grid approach that scans over a predefined set of DMA analog combining matrices, focusing on certain 3D locations in the near-field regime, to find the one unveiling the UE 3D position. In particular, we deploy the predefined set of $\W_{\rm RX}$ to obtain a set of quantized received pilot signals. At every TTI, the DMA RX receives with each of the available combining matrices, aiming to effectively scan its designated vicinity, and updates $k$ in \eqref{eq: thresh} accordingly. Apparently, the closer the true UE location to a focusing analog combiner, the larger is the captured pilot signal strength by the considered DMA RX, whereas, when focusing away than that position (in any dimension), the received signal is weakened. To this end, by deploying the quantizer described in \eqref{eq: quan}, we can distinguish between environmental noise and transmitted pilot symbols. In fact, when the combiner focuses closer to the UE position, a higher number of quantized received signals is recognized as pilot signals, allowing us to generate a pseudo spectrum indicating the effectiveness of each analog combining phase profile $\W_{\rm RX}$.

By searching over a predefined set of potential UE positions $(\widehat{r}_p,\widehat{\theta}_p,\widehat{\varphi}_p)\in \mathcal{S}$ with $p=1,2,\ldots,|S|$, we systematically construct the virtual channel $\widehat{\h}$ for each position. This search centers around a virtual UE located at the coordinates $(\widehat{r},\widehat{\theta},\widehat{\varphi})$. We derive the DMA analog combining matrix $\W_{\rm RX}$ by solving $\mathcal{OP}$, and by observing the resulting $\y_q$, we build a pseudo-spectrum $\mathcal{P}$ to evaluate the effectiveness of each $\W_{\rm RX}$ configuration. After iterating over all possible entries of $\mathcal{S}$, and consequently, conducting a grid search on the pseudo-spectrum $\mathcal{P}$, we can accurately determine the UE's 3D position. This procedure is summarized in Algorithm~1.

\section{Numerical Results and Discussion}\label{sec: num}
\begin{figure}[!t]
\centering
\includegraphics[width=\columnwidth]{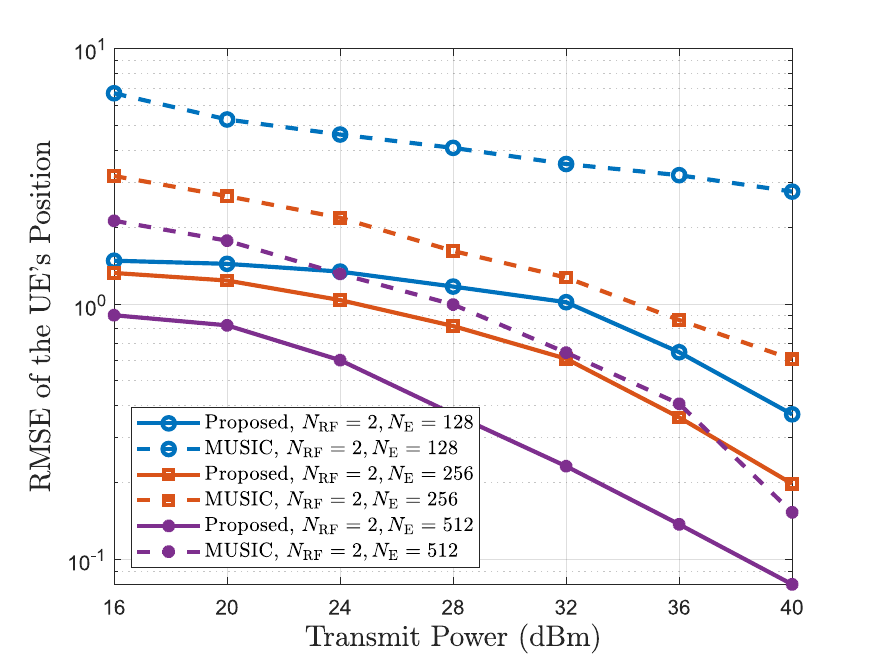}\vspace{-0.2cm}
\caption{\small{RMSE performance of the UE position estimation versus the transmit power $P_{\rm max}$ in dBm for a DMA-based hybrid A/D RX with $N_{\rm RF}=2$ RF chains, $N_{\rm E}=\{128,256,512\}$ metamaterials, and $|S|=T=200$ UE grid positions, and thus, analog combiners.}
}\vspace{-0.4cm}
\label{fig:RMSE}
\end{figure}
In this section, we evaluate the performance of the proposed near-field localization framework for DMA-based hybrid A/D RXs with $1$-bit resolution ADCs. We focus on the sub-THz central frequency of $140$ GHz with a $150$ KHz bandwidth, and consider extremely large DMAs with inter-microstrip and inter-element distances $d_{\rm RF}=\lambda/2$ and $d_{\rm E}=\lambda/5$, respectively. The single-antenna UE was assumed randomly positioned with spherical coordinates $\phi_u = 90^{\circ}$, $\theta_u \in [0^{\circ}, 90^{\circ}]$, and radial distance $r_u \in [1, 20]$ meters, ensuring placement within the Fresnel region. Given the inevitable dependence of the proposed localization approach on $|S|$, we have made the assumption that prior knowledge of the UE's coordinates is available. To this end, the DMA analog combining focused on the confidence interval $[d_\ell-\ell, d_\ell+\ell]$, where string $\ell\in\{r, \theta\}$ represents the true value of the coordinate and $d_\ell$ signifies the width of the interval. The confidence interval was considered inside the set of the latter valid UE coordinates. The noise variance $\sigma^2$ was set to $-174 + 10\log_{10}(B)$ in dB and a $10$-bit beam codebook $\mathcal{F}$ was used for the DMA analog combiners $\W_{\rm RX}$. All performance evaluations were averaged over $300$ independent Monte Carlo realizations.

\begin{figure}[!t]
\centering
\includegraphics[width=\columnwidth]{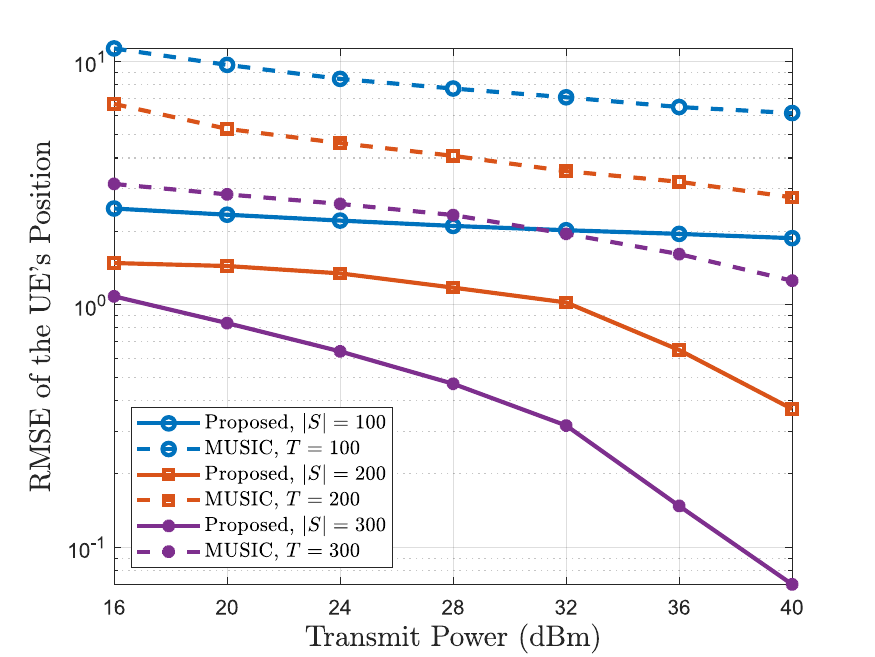}\vspace{-0.2cm}
\caption{\small{RMSE performance of the UE position estimation versus the transmit power $P_{\rm max}$ in dBm for a DMA-based hybrid A/D RX with $N_{\rm RF}=2$ RF chains, $N_{\rm E}=128$ metamaterials, and $|S|=T=\{100,200,300\}$ UE grid positions, and thus, DMA analog combiners.}
}\vspace{-0.4cm}
\label{fig:Res}
\end{figure}

In Figs.~\ref{fig:RMSE} and~\ref{fig:Res}, we illustrate the RMSE of the proposed localization approach in Algorithm~\ref{alg:loc} for different number of metamaterials $N_{\rm E}$, grid searching overhead $T$, and transmit power levels $P_{\rm max}$ in dBm. As a baseline, we have include in both figures the performance of the MUSIC-variant presented in \cite{gavras2023full} for full resolution ADCs at the RX RF chains. In both figures, we have set $d_r=5$ meters and $d_{\theta}=10^{\circ}$. As shown, and as expected, the localization performance improves with increasing SNR. In Fig.~\ref{fig:RMSE}, considering an extremely large DMA with $N_{\rm E} = \{128,256,512\}$ metamaterials and $N_{\rm RF}=2$ RF chains, and $|S|=T=200$ TTIs, it can observed that the proposed scheme outperforms \cite{gavras2023full}'s localization framework for increasing $N_{\rm E}$ values. In Fig.~\ref{fig:Res}, $N_{\rm RF}=2$ RX RF chains and $N_{\rm E} = 128$ metamaterials per microstrip were considered as well as $|S|=T=\{100,200,300\}$ TTI values. As shown, the proposed algorithm outperforms the baseline even with a limited number of TTIs, indicating that its scalability is compromised with fewer TTIs. This limitation is reasonable because reducing $|S|$ leads to a sparser sampling of points in space, resulting in a less accurate estimation. On the contrary, with a reduced number $N_{\rm E}$ of metamaterials, the algorithm can still achieve satisfactory estimation performance if there is large $|S|$. This observation underscores the importance of having an ample overhead to compensate for limitations sourced on $N_{\rm E}$, demonstrating the trade-off between the estimation overhead $T$ and number $N_{\rm E}$ of DMA metamaterials for achieving accurate estimations.

\section{Conclusion}
In this paper, we presented a near-field localization framework with a DMA-based hybrid A/D RX architecture, including $1$-bit resolution ADCs, one at each of its reception RF chains that is fed by a distinct microstrip. The proposed algorithm includes a grid search over predefined analog combiners at the DMA, which represent candidate UE positions in the near field. It was demonstrated that the proposed scheme outperforms a baseline approach with full resolution ADCs, especially for DMAs with large sizes of metamaterials per microstrip.

\newpage
\bibliographystyle{IEEEtran}
\bibliography{IEEEabrv,ms}
\end{document}

%% file: Definitions.tex
\newcommand{\g}{\mathbf{g}}
\newcommand{\h}{\mathbf{h}}

\newcommand{\n}{\mathbf{n}}

\newcommand{\y}{\mathbf{y}}

\renewcommand{\P}{\mathbf{P}}

\newcommand{\W}{\mathbf{W}}

\newcommand{\Compl}{\mbox{$\mathbb{C}$}}
